# The Product Life Cycle of Durable Goods

**Joachim Kaldasch[1*]**

*[1]EBC Hochschule Berlin, Alexanderplatz 1, 10178 Berlin, Germany.*

***Author's contribution***

*The sole author designed, analyzed and interpreted and prepared the manuscript.*

***Article Information***

DOI: 10.9734/BJEMT/2015/20395
*Editor(s):*
(1) Chen Zhan-Ming, School of Economics, Renmin University of China, Beijing, China.
*Reviewers:*
(1) Anonymous, Shinawatra University, Thailand.
(2) Anonymous, Shou University, Taiwan
Complete Peer review History: http://sciencedomain.org/review-history/11281

***Original Research Article***

***Received 25th July 2015***
***Accepted 25th August 2015***
***Published 6th September 2015***

## ABSTRACT

A dynamic model of the product lifecycle of (nearly) homogeneous durables in polypoly markets is established. It describes the concurrent evolution of the unit sales and price of durable goods. The theory is based on the idea that the sales dynamics is determined by a meeting process of demanded with supplied product units. Taking advantage from the Bass model for first purchase and a logistic model for repurchase the entire product lifecycle of a durable can be established. For the case of a fast growing supply the model suggests that the mean price of the good decreases according to a logistic law. Both, the established unit sales and price evolution are in agreement with the empirical data studied in this paper.

The presented approach discusses further the interference of the diffusion process with the supply dynamics. The model predicts the occurrence of lost sales in the initial stages of the lifecycle due to supply constraints. They are the origin for a retarded market penetration. The theory suggests that the imitation rate *B* indicating social contagion in the Bass model has its maximum magnitude for the case of a large amount of available units at introduction and a fast output increase. The empirical data of the investigated samples are in qualitative agreement with this prediction.

**Corresponding author: E-mail: joachim.kaldasch@international-business-school.de;*

## 1. INTRODUCTION

A microeconomic model is presented that investigates the product lifecycle (PLC) of (nearly) homogeneous durables in polypoly markets.[1] The product lifecycle concept suggests that analogous to the life of organisms goods are subject to characteristic stages in their unit sales evolution classified usually into introduction, growth, maturity and decline phase [1]. The PLC of durables was studied intensively in order to forecast the adopter and sales evolution [2-4]. It also applies as a guideline for corporate marketing strategy and in operation research in order to estimate production capacities [5-7].

The unit sales evolution is determined by first- and repurchase of a durable. First purchase is related to the spreading of a good into the market. Repurchase is proportional to the number of current adopters and a corresponding repurchase rate that describes multiple and replacement purchase of product units. The evolution of the market penetration is known as market diffusion stipulating the initial stages of the PLC. In particular the Bass model turned out to give an appropriate picture of the diffusion process of consumer durables [8]. It suggests that the market diffusion of new products consists of a fast and a slow spreading wave. The fast wave is mediated by mass media and the slow one is due to social contagion. While diffusion models work well as a forecast tool the implementation of the price as a decision variable turned out to be difficult. A variety of models based on the Bass model were developed in order to include the price. It is usually taken into account as a perturbation of the spreading process [4,9,10]. The price, though, is a key microeconomic variable that has its origin in the relation between supply and demand. The presented dynamic model offers an alternative approach to understand the unit sales and price evolution not only of durables but also of non-durable goods in a unique framework [11].

The key idea to combine the sales and price dynamics of a durable is to treat purchase events as a meeting process of demanded (required) with supplied (available) product units. While the spreading process determines the generation rate of demanded units, the number of available units is a consequence of the output flow of suppliers. It can be shown that based on this idea the price dispersion of homogeneous goods has for short time periods the form of a Laplace distribution [12,13]. The mean price of this distribution is determined by a Walrus equation. It suggests that an excess growth of demanded units cause an increase and an excess growth of supplied units a decrease of the mean price [14]. The presented model is an application of these considerations and rigorously derives the PLC of durable goods from the dynamics of demand and supply.

The paper is organized as follows. The next section establishes a model for the PLC of durables in polypoly markets. It starts with the derivation of the market dynamics of demanded and supplied product units followed by a consideration of the price dynamics of homogenous goods. After deriving the sales and price dynamics of durables the impact of the supply evolution on the rate of adoption is studied. The presented approach is applied to empirical investigations of the PLC of nearly homogenous consumer durables where market penetration and price data are available. The theory of the PLC of durable goods is summarized in the discussion, followed by the main conclusions of this work.

## 2. THE MODEL

### 2.1 The Dynamics of Polypoly Markets

The theory presented here is established for (nearly) homogeneous durable goods in polypoly markets. The demand side of the durable market is determined by the total number of demanded (desired) units $\tilde{x}(t)$ at time step $t$ generated by potential buyers. The supply side on the other hand is given by the total number of supplied (available) units $\tilde{z}(t)$ offered by $N(t)$ suppliers (retailers). For a polypoly market we demand that $N(t) \gg 1$.

The total unit sales of the durable are indicated $\tilde{y}(t)$. As mentioned above the main idea of the presented microeconomic approach is to consider purchase events as the meeting of demanded with supplied product units. Therefore $\tilde{y}(t)$ must vanish if either $\tilde{x}(t)$ or $\tilde{z}(t)$ disappears. Hence the total unit sales can be

[1] *With nearly homogeneous goods is meant that the good cannot be treated as comprising of successive technological generations. In this case we would have to treat each generation separately as homogeneous. For example not the personal computer but each technological generation of personal computers can be regarded as homogeneous.*

written up to the first order as a product of both variables [15][2]:

$$\tilde{y}(t) \cong \eta \tilde{z}(t)\tilde{x}(t) \quad (1)$$

where the unknown rate $\eta$ characterizes the mean frequency by which the meeting process generates successful purchase events. Since $\tilde{x}(t)$, $\tilde{z}(t)$, $\tilde{y}(t) \geq 0$, we demand that also $\eta \geq 0$. The evolution of the number of demanded and supplied units can be written as conservation relations of the form[3]:

$$\frac{d\tilde{x}(t)}{dt} = \tilde{d}(t) - \tilde{y}(t) \quad (2)$$

and

$$\frac{d\tilde{z}(t)}{dt} = \tilde{s}(t) - \tilde{y}(t) \quad (3)$$

Eq. (2) suggests that the total number of demanded units increases with the total demand rate $\tilde{d}(t)$ which represents the generation rate of demanded units by potential buyers. The number $\tilde{x}(t)$ decreases in time by the purchase of product units with the total unit sales rate $\tilde{y}(t)$. Eq. (3) states that the total number of supplied units increases by the supply of product units with the total supply rate (total output) $\tilde{s}(t)$ and decreases due to the purchase of these units with the total unit sales rate $\tilde{y}(t)$. The variables in Eq. (3) can be obtained in a polypoly market from a sum over the number of suppliers $N(t)$ by:

$$\tilde{z}(t) = \sum_{i=1}^{N(t)} z_i(t); \quad \tilde{s}(t) = \sum_{i=1}^{N(t)} s_i(t); \quad \tilde{y}(t) = \sum_{i=1}^{N(t)} y_i(t) \quad (4)$$

#### 2.1.1 Demand and supply

In order to establish the market evolution of a durable the dynamics of demand and supply has to be specified. We assume that demanded units do not last forever ones they are generated. We take into account that they have a finite mean lifetime $\Theta$. That means, demanded units not leading to purchase events during a time period $\Theta$ disappear. This effect can be included in the demand rate $\tilde{d}(t)$ by writing:

$$\tilde{d}(t) = \tilde{d}_0(t) - \frac{\tilde{x}(t)}{\Theta} \quad (5)$$

where $\tilde{d}_0(t)$ describes the generation rate of demanded units by potential buyers and the second term takes the disappearance of demanded units with the rate $1/\Theta$ into account. The amount of demanded units generated by the rate $\tilde{d}_0(t)$ can be given by:

$$\tilde{x}_0(t) = \Theta \tilde{d}_0(t) \quad (6)$$

The supply side is governed by the reproduction process. In a free market suppliers sell product units in order to make profit. Reinvesting a part of their profit and external money they can increase the total output $\tilde{s}(t)$ in time. This growth process can be characterized by the variable $\gamma(t)$, defining the relation between total supply flow and total unit sales: [4]

$$\gamma(t) = \frac{\tilde{s}(t)}{\tilde{y}(t)} - 1 \quad (7)$$

With this relation Eq. (3) turns into[5]:

$$\frac{d\tilde{z}(t)}{dt} = \gamma(t)\tilde{y}(t) \quad (8)$$

We want to confine here to fast growing markets with $\gamma(t)>0$. For a sufficiently high $\gamma$ the number of supplied units evolves much faster than the number of demanded units in the considered time interval $\Delta t$ such that:

$$\frac{d\tilde{x}(t)}{dt} << \frac{d\tilde{z}(t)}{dt} \quad (9)$$

---

*[2] Note that this form of relation has its origin in the description of chemical reactions in terms of rate equations, where the reaction rate is proportional to the product of the concentrations of the reactants. This form was applied also to other meeting processes in physics, chemistry, biology and economy. So for example the second term of the Bass model considered below has a similar form describing the meeting of potential adopters with current adopters. Eq.(1) is not applicable to the case of rare purchase events known for example from luxury goods.*

*[3] In order to establish a continuous model integer variables are scaled by a large constant figure such that they can be treated as small real numbers. We demand that this scaling leads to* $\tilde{x}(t), \tilde{z}(t) \leq 1$.

*[4] This variable is also called reproduction parameter, since it characterizes the growth process of the output in the reproduction process.*

*[5] The impact of the finite lifetime of a durable is neglected because it is large compared to the case of non-durables (see [11]).*

In this case we can approximate:[6]

$$d\tilde{x}(t)/dt \cong 0 \tag{10}$$

and obtain immediately from Eq. (2) and Eq. (5):

$$\tilde{y}(t) \cong \tilde{d}(t) = \tilde{d}_0(t) - \frac{1}{\Theta}\tilde{x}(t) \tag{11}$$

In this approximation the unit sales are equal to the generation rate of demanded units diminished by the rate $\tilde{x}(t)/\Theta$. Applying Eq.(1) and Eq. (6) in Eq. (11) we get for the total number of demanded units:

$$\tilde{x}(t) = \frac{\tilde{x}_0(t)}{1+\Theta\eta\tilde{z}(t)} \tag{12}$$

Expanding this relation for small $\tilde{z}(t)$ yields:

$$\tilde{x}(t) \cong \tilde{x}_0(t)\left(1-\Theta\eta\tilde{z}(t)\right) \tag{13}$$

In order to determine the time evolution of $\tilde{z}(t)$ we apply Eq. (13) in Eq. (1). Eq. (8) turns into:

$$\frac{d\tilde{z}(t)}{dt} = \alpha(t)\tilde{z}(t) - \Theta\eta\alpha(t)\tilde{z}(t)^2 \tag{14}$$

with:

$$\alpha(t) = \eta\gamma\tilde{x}_0(t) \tag{15}$$

In order to solve Eq. (14) we approximate the time dependent growth rate of supplied units *α(t)* by its time average over the considered time interval *Δt*:

$$\alpha = \frac{1}{\Delta t}\int_{t_0}^{t_0+\Delta t}\alpha(t)dt \tag{16}$$

In this approximation Eq. (14) becomes a logistic differential equation with constant coefficients. The evolution of the number of supplied units can be given by:

$$\tilde{z}(t) = \frac{z_{\max}}{1+C_z e^{-\alpha t}} \tag{17}$$

[6] *This approach is known as adiabatic approximation.*

with the integration constant $C_z$ and:

$$z_{\max} = \frac{1}{\eta\Theta} \tag{18}$$

The time evolution of the total number of available units depends on the sign of the supply growth rate *α*. For the considered supply market with *α>0*, Eq. (17) predicts that $\tilde{z}(t)$ increases in time according to a logistic law until $z(t)=z_{max}$. Inserting $z_{max}$ in Eq. (12) suggests that the number of demanded units becomes $\tilde{x}(t) = \tilde{x}_0(t)/2$ in the saturated state. This consideration is based on the condition that no supply constraints occur during *Δt*.[7]

The total unit sales Eq. (11) have with Eq. (12) and Eq. (17) the form:

$$\tilde{y}(t) = \tilde{d}_0(t)\frac{z(t)}{z_{\max}} = \frac{\tilde{d}_0(t)}{1+C_z e^{-\alpha t}} \tag{19}$$

This relation suggests that the sales evolution is determined on the one hand by the generation rate of demanded units $\tilde{d}_0(t)$ and on the other hand by the evolution of available units $\tilde{z}(t)$. At introduction of the good $\tilde{z}(t)$ is necessarily small. Therefore the unit sales might be limited by the amount of available units leading to so-called lost sales.[8] With an increasing supply flow this effect disappears.

### 2.1.2 The mean price evolution of homogeneous goods

In order to take the product price *p* of the homogeneous good as a decision variable of potential buyers into account, we introduce price dependent cumulated distribution functions of demanded and supplied units *x(p,t)* and *z(p,t)* by:

$$x(t,p) = \tilde{x}(t)\int_0^p x'(t,p')dp' \tag{20}$$

$$z(t,p) = \tilde{z}(t)\int_0^p z'(t,p')dp' \tag{21}$$

[7] *If there are supply constrains the number of supplied units is bounded by a $z_{max}$ that is smaller than given by Eq.(18). In this case the supplied units evolve in logistic waves found in empirical investigations of durables discussed below.*

[8] *This effect is also termed backordering.*

where *x'(p,t)* and *z'(p,t)* are the probability densities to find demanded respectively supplied units in a given price interval *p* and *p+dp*. The functions *x(p,t)* and *z(p,t)* can be interpreted as demand and supply curves of the good. Generalizing Eq. (1) we assume that the number of sold units in a given price interval must disappear if the corresponding numbers of *x(t,p)* respectively *z(t,p)* vanish. Hence the price dependent unit sales are up to the first order proportional to both variables:

$$y(t,p) \cong \eta x(t,p) z(t,p) \tag{22}$$

where the meeting rate $\eta$ is treated as price independent. The price dispersion of sold units of a homogeneous good is determined by the probability density $P_y(t,p)$ defined by:

$$P_y(t,p) = \frac{y(t,p)}{\tilde{y}(t)} \tag{23}$$

As established in [12,13] the price dispersion of homogeneous goods can be derived from Eq. (22). For short time horizons the price dispersion has the form of a symmetric Laplace distribution:

$$P_y(p) \cong \frac{1}{2\sigma} e^{-\frac{|p-\mu|}{\sigma}} \tag{24}$$

with the standard deviation:

$$Std\left(P_y(p)\right) = \sqrt{2}\sigma \cong \sqrt{2}\left(\mu - \mu_m\right) \tag{25}$$

where $\mu$ is the mean price of the distribution and $\mu_m>0$ is a minimum mean price indicating a technological limit beyond which the supply of product units is not profitable. Further derived in these references is a relation that governs the dynamics of the mean price $\mu(t)$. It suggests that the mean price is determined by a Walrus equation of the form:

$$\frac{1}{\mu - \mu_m}\frac{d\mu}{dt} = H\left(\frac{d\tilde{x}}{dt} - \frac{d\tilde{z}}{dt}\right) \tag{26}$$

where $H>0$ is a constant. This relation can be used to characterize the evolution of the mean price of homogeneous durables in the considered market constellation. For this purpose we take advantage from Eq. (10) and rewrite Eq. (26) as:

$$\frac{1}{\mu(t) - \mu_m}\frac{d\mu(t)}{dt} \cong -H\frac{d\tilde{z}(t)}{dt} \tag{27}$$

That means, in a fast growing polypoly market the mean price is essentially determined by the evolution of available product units. Applying Eq. (14) we get:

$$\frac{d\mu(t)}{dt} \cong -H\alpha\tilde{z}(t)\left(\mu(t) - \mu_m\right) \tag{28}$$

while higher order terms in $\tilde{z}(t)$ are neglected. For $\alpha>0$ the mean price of a durable is declining in time due to the excess supply.[9] The stationary solution of this relation is given either by $\mu=\mu_m$ or $\tilde{z} = z_{\max}$ . Eq. (25) suggests that for $\mu=\mu_m$ the standard deviation disappears and market becomes a monopoly market. Since we focus here on polypoly markets this case is not further considered.

For $\mu(t)>\mu_m$, Eq. (27) can be written as:

$$\int \frac{d\mu(t)}{\mu(t) - \mu_m} \cong -H\int d\tilde{z}(t) \tag{29}$$

and we readily obtain:

$$\mu(t) = \mu_0 e^{-H\tilde{z}(t)} + \mu_m \tag{30}$$

The model suggests that for a polypoly market the mean price declines with increasing $\tilde{z}(t)$ given by Eq. (17). For $\tilde{z}(t) \to z_{\max}$ the mean price approaches a floor price $\mu_f>\mu_m$ governed by a logistic law. The floor price can be obtained from:

$$\mu_f = \mu_0 \exp(-Hz_{\max}) + \mu_m \tag{31}$$

The introduction mean price of the good $\mu(0)$ is defined by:

$$\mu(0) = \mu_0 \exp(-H\tilde{z}(0)) + \mu_m \tag{32}$$

while generally $\tilde{z}(0) \neq 0$ .

## 2.2 The Unit Sales Evolution

The evolution of the total unit sales of a good is usually referred to as product lifecycle. The total unit sales are the result of the first- and repurchase of the durable. First purchase is related to the spreading of the good into the market. This spreading process can be described by the cumulative number of adopters $N_A(t)$. In order to be in line with previous research we want to define the market penetration $n(t)$ by[10]:

$$n(t) = \frac{N_A(t)}{M} \quad (33)$$

[9] *For $\alpha<0$ the mean price would exhibit an exponential growth caused by a supply shortage.*

The number of all possible adopters in a given domain interested in purchasing the consumer durable is termed market potential $M$.[11]

We want to take the inhomogeneity of the demand side of the market with respect to the price into account by introducing a market volume $V(\mu) \leq M$. It represents the number potential adopters who can afford the good for a mean price $\mu(t)$. The scaled market volume becomes:

$$v(\mu) = \frac{V(\mu(t))}{M} \quad (34)$$

while $0 \leq v(\mu) \leq 1$. The market volume decreases with increasing mean price, $dv(\mu)/d\mu < 0$.

First purchase is determined by the time evolution of the number of adopters. This number is governed by a conservation relation and has the form:

$$\frac{dn(t)}{dt} = \varphi(t)\psi(t) - \theta n(t) \quad (35)$$

The first term is the product of a generation rate $\varphi(t)$ of adopters with the number of potential adopter $\psi(t)$ not yet adopted the good. The variable $\psi(t)$ can be written as the difference between the market volume and the current number of adopters:

$$\psi(t) = v(\mu) - n(t) \quad (36)$$

The second term in Eq. (35) indicates the decline of $n(t)$ with a mean decline rate $\theta$. In the initial stages of the lifecycle is $\varphi(t)>0$, while $\theta \approx 0$. The decline phase is characterized by $\varphi(t) \approx 0$, $\theta>0$. Expanding the generation rate $\varphi(t)$ as a function of the number of adopters leads to:

$$\varphi(t) \cong \mathrm{A} + \mathrm{B}n(t) \quad (37)$$

with constant coefficients $A, B>0$. Inserting Eq. (37) in Eq. (35) with $\theta=0$, we obtain a standard model for the spreading process of goods known as the Bass model [16]. The first purchase unit sales are determined by the growth of the market penetration:

$$\tilde{y}_f(t) = \frac{dn(t)}{dt} = \mathrm{A}(v(\mu) - n(t)) + \mathrm{B}n(t)(v(\mu) - n(t)) \quad (38)$$

The Bass model suggests that the spreading of a good into a market consists of two waves. The first term indicates the fast wave due to the spontaneous purchase of potential adopters with the so-called innovation rate $A$. The second term indicates a slow spreading wave caused by social learning, where the number of adopters $n(t)$ increases with an imitation rate $B$.

Repurchase events must be proportional to the current number of adopters. The total repurchased number of units per unit time $\tilde{y}_r(t)$ can therefore be modelled as the product of $n(t)$ with a time dependent repurchase rate $\xi(t)$ characterizing the average number of repurchased units per unit time and adopter:

$$\tilde{y}_r(t) = \xi(t)n(t) \quad (39)$$

The repurchase rate $\xi(t)$ describes multiple and replacement purchase. Replacement purchase is due to the finite lifetime $\tau$ of the good caused for example by product failure. Multiple purchases can be specified by the number of units $\iota$ purchased during this mean lifetime. The repurchase rate $\xi(t)$ approaches therefore after sufficient time to a maximum magnitude $\xi_{max}=\iota/\tau$.[12] The growth of the repurchase rate up to $\xi_{max}$ can be modelled by the dynamic equation:

$$\frac{d\xi(t)}{dt} = a\xi(t)(\xi_{\max} - \xi(t)) \quad (40)$$

[10] *As mentioned above integer numbers are scaled by a large number. In order to establish a consistent model this large number is M.*

[11] *For simplicity M is treated as time independent.*

[12] *Note that virtual goods like music, software or books have in infinite lifetime $\tau \to \infty$ and hence $\xi_{max} \approx 0$. Therefore repurchase of these goods usually disappear.*

where $a>0$ characterizes the growth rate of $\xi(t)$. The solution of this relation determines the repurchase rate evolution:

$$\xi(t)=\frac{\xi_{\max}}{1+C_{\xi}e^{-at}} \tag{41}$$

with the free parameter $C_{\xi}$.

The total unit sales are the sum of first- and repurchase events:

$$\tilde{y}(t)=\tilde{y}_f(t)+\tilde{y}_r(t)=\frac{dn(t)}{dt}+\xi(t)n(t) \tag{42}$$

The product lifecycle can be characterized by the main processes governing the adopter evolution:

i) In the introduction phase of the PLC the main source of the growth of $n(t)$ is spontaneous purchase. Hence, the introduction phase is determined by the dominance of the first term in Eq. (38) over the second which leads to the condition $A>Bn(t)$. It is related to the fast spreading wave of the good.
ii) In the growth phase social learning dominates the adopter evolution. This phase is related to the slow spreading wave where $A<Bn(t)$ in Eq. (38). At the transition between both phases the number of adopters is $n_g \approx A/B$.
iii) In the maturity phase the spreading due to social contagion saturates while:

$$\frac{dn(t)}{dt}\approx 0 \tag{43}$$

From Eq. (38) follows with Eq. (43) that the number of adopters is determined in this period by:

$$n(t)\cong v(\mu(t)) \tag{44}$$

However, according to Eq. (30) the mean price decreases in time and as a consequence the market volume $v(\mu(t))$ expand. That means, when the spreading process due to social contagion slows down the number of adopters may increase induced by the decline of the mean price. However, in order to keep the presented model simple this second spreading wave in the maturity phase of the PLC is not further taken into account here. [13] The market volume is approximated here by a constant $v(\mu)=n_0$. The evolution of the market penetration $n(t)$ can then be described by the solution of Eq. (38) given by:

$$n(t)=\frac{1-e^{-(A+B)t}}{\left(1+\frac{B}{A}e^{-(A+B)t}\right)^2}n_0 \tag{45}$$

which yields an $S$-curve of the market penetration in time. The total first purchase unit sales caused by Bass diffusion become:

$$\tilde{y}_f(t)=\frac{dn(t)}{dt}=\frac{A(A+B)^2e^{-(A+B)t}}{\left(A+Be^{-(A+B)t}\right)^2}n_0 \tag{46}$$

iv) The decline phase is related to the occurrence of a close substitute of the current durable at time step $t_d$. This substitute changes the repurchase behaviour of potential buyers immediately. For the case that the current good is ultimately replaced the repurchase rate $\xi(t)$ declines in time while $\xi_{max}$ vanishes. This effect can be taken into account in the time evolution of the repurchase rate Eq. (40) by assuming a negative growth rate $-a'$. Eq. (40) turns therefore for $t>t_d$ into:

$$\frac{d\xi(t)}{dt}\cong -a'\xi(t) \tag{47}$$

with $a'>0$ and neglecting higher order terms in $\xi(t)$. The differential equation has the solution:

$$\xi(t)=\xi(t_d)e^{-a'(t-t_d)}+\xi_{\min} \tag{48}$$

where $\xi_{min}$ indicates a remaining minimum repurchase rate. As a result of the declining repurchase rate the number of adopters declines with the rate $\theta>0$. Applying Eq. (35) with $\varphi(t>t_d)=0$ we get for $t>t_d$:

$$n(t)=n(t_d)e^{-\theta(t-t_d)}+n_{\min} \tag{49}$$

while $n_{min}$ is an eventually remaining minimum number of adopters.

[13] *Models including the price evolution are discussed for example in [10,15].*

The established relations allow a characterization of the entire product lifecycle for both, the evolution of the mean price and the unit sales and penetration evolution. In order to compare the model with empirical data a minimum set of free parameters is required. The evolution of $\tilde{z}(t)$ entails three free parameters $z_{max}$, $\alpha$ and $C_z$. For the mean price evolution four additional parameters are necessary $\mu_0$, $\mu_f$, $\mu_m$ and $H$. The adopter evolution is characterized here by Bass diffusion Eq. (45). It involves the free parameters $A$, $B$ and $n_0$. The decline phase starts at $t_d$ and requires two extra parameters $\theta$ and $n_{min}$. The total unit sales $\tilde{y}(t)$ are governed by Eq. (42) determined by Bass diffusion and the repurchase rate parameters $a$, $\xi_{max}$ and $C_\xi$. The decline phase finally uses the parameters $a'$ and $\xi_{min}$. Since the empirical unit sales are usually given in absolute numbers, for a comparison the market potential $M$ has to be known. A complete description of the product lifecycle together with the mean price evolution involves a minimum number of 19 free parameters in the presented approximation.

## 2.3 The Relation between Market Penetration and Supply

Eq. (19) suggests that the initial stages of the lifecycle suffer from a constrained supply. This effect must have an impact on the velocity of the spreading process of a durable. Here we want to estimate the impact of a constrained supply on the imitation parameter $B$ of the Bass model. For this purpose the total unit sales are approximated by the first purchase sales $\tilde{y}_f(t)$ in Eq. (42). Further the impact of spontaneous purchase on $\tilde{y}_f(t)$ can be neglected since A<<1. From Eq. (19) follows with Eq. (38):

$$\mathrm{B}n(t)\psi(t) \cong \frac{\tilde{d}_0(t)}{1+C_z e^{-\alpha t}} \tag{50}$$

Obviously the impact of a constraint supply can be neglected if $C_z=0$. The idea is therefore to treat the imitation parameter $B$ as a function of $C_z$. Expanding $B(C_z)$ up to the first order yields:

$$\mathrm{B}(C_z) \cong B_0 + B_1 C_z \tag{51}$$

with the unknown coefficients $B_0$ and $B_1$. Inserting this relation in Eq. (50) and expanding also the right hand side of Eq. (19) with respect to $C_z$ we get:

$$\left(B_0 + B_1 C_z\right) n(t)\psi(t) \cong \tilde{d}_0(t)\left(1 - C_z e^{-\alpha t}\right) \tag{52}$$

A comparison of the coefficients leads to:

$$B_0 n(t)\psi(t) = \tilde{d}_0(t); \qquad B_1 n(t)\psi(t) = -\tilde{d}_0(t)e^{-\alpha t} \tag{53}$$

The relations suggest that $B_1$ must be a time dependent function of the form:

$$B_1(t) = -B_0 e^{-\alpha t} \tag{54}$$

Taking the time average over the growth period $\Delta t_g$, we obtain for the mean magnitude of the imitation parameter:

$$\mathrm{B}(\beta) = \frac{B_0}{\Delta t_g}\int_0^{\Delta t_g}\left(1 - C_z e^{-\alpha t}\right)dt = B_0\left(1+\beta\right) \tag{55}$$

with:

$$\beta = \frac{e^{-\alpha \Delta t_g} - 1}{\alpha \Delta t_g} C_z \tag{56}$$

Note that $\beta \leq 0$, because the exponential function is smaller than one for $\Delta t_g > 0$. Eq. (55) suggests that $B(\beta)$ is linear increasing function. The magnitude of the imitation parameter rises with increasing $\beta$ until $\beta=0$. For $\beta=0$, the supply constraint has no impact on the imitation parameter. In this case $B(\beta)$ has its maximum magnitude $B_0$ and the diffusion process takes place with maximum velocity. The diffusion is unperturbed for the case of a fast price decline related to a high $\alpha$ during the growth phase $\Delta t_g$, respectively for small $C_z$ indicating a large amount of available units at introduction of the good. It has to be emphasized that the retarded market diffusion is not a consequence of the price evolution but exclusively the result of the constrained supply in the initial stages of the lifecycle.

## 3. COMPARISON WITH EMPIRICAL RESULTS

The presented model suggests that the PLC of (nearly) homogeneous consumer durables in polypoly markets have the following characteristics:

1. Homogeneous durable goods in polypoly markets have a price dispersion that can be approximated by the Laplace distribution Eq. (24).
2. For a fast growing supply the number of available units $\tilde{z}(t)$ increases in time according to the logistic law Eq. (17). This evolution causes a logistic decline of the mean price described by Eq. (30).
3. Applying the Bass model for the market penetration $n(t)$ the initial stages of the PLC can be approximated by Eq. (45), while the decline phase is characterized by Eq. (49).
4. The total unit sales are determined on the one hand by the growth of the number of adopters and on the other hand by the repurchase of the good. First purchases are given in the presented approximation by Eq. (46) as a result of the Bass model. Repurchase is determined by Eq. (39) with the logistic repurchase rate Eq. (41). The decline phase is governed by an exponential decay of the repurchase rate described by Eq. (48).
5. The diffusion processes is subject to supply constraints causing a decelerated diffusion process. Plotting the imitation rate $B$ as a function of $\beta$ the model suggests a linear relationship of the form Eq. (55).

In order to compare the presented model with available empirical data a number of consumer durables are studied satisfying the model conditions where the market penetration and the price evolution are known. For two samples also data of the empirical unit sales are available. Unfortunately the first assertion cannot be verified for the considered samples because the price dispersions of these goods are unknown. From empirical data of the price dispersion of homogeneous goods it is known, however, that their price dispersion can be approximated by a Laplace distribution [13,16,17]. It is assumed that the chosen examples exhibit similar distributions.

### 3.1 Market Penetration, Mean Price and Supply Evolution

In order to illustrate the entire PLC of a durable, the empirical market penetration $n(t)$ of Black & White (B&W) TV sets in the UK [18] is displayed in Fig. 1 (squares) together with a fit of the Bass model Eq. (45) for the initial stages of the PLC (solid line) while Eq. (49) is applied for the decline phase. The fit parameters are summarized in Table 1. Indicated are the phases characterizing the stages of the lifecycle as suggested by the model. The transition between introduction and growth phase takes place at a market penetration $n_g \approx 2\%$. We assume that deviations from the Bass model in the maturity phase of the PLC are due to an additional diffusion process not further discussed here. The decline phase of this good can be described by an exponential decay of the number of adopters after $t_d=26$ years suggested by Eq. (49). Displayed in Fig. 1 as a measure of the mean price $\mu(t)$ is the average sales price scaled by the maximum price [19]. The dashed line represents a fit of Eq. (30). Shown in the insert is the function $\tilde{z}(t)$ that follows from the fit of the mean price with the scaling $z_{max}=1$. According to this model $\tilde{z}(t)$ indicates the evolution of freely available units. Shown in Fig. 2 is a similar example. Displayed is the empirical unit sales evolution of Black & White TV sets in the USA [19]. Both samples have a dynamics in agreement with the presented model applying the parameters summarized in Table 1. Unfortunately data for the decline phase of this durable are not available.

Presented in Fig. 3 are the empirical market penetration and price evolution of CD-players in the USA [20]. This good has the smallest parameter $\alpha=0.17$ in the considered collection of durables suggesting a slow increase of the total output in time. A consequence of the theory is that for small supply growth rates $\alpha$ the mean price is susceptible for demand and supply fluctuations. This can be seen in the initial period of the market penetration where a considerable fluctuation of the mean price is evident. Approaching the inflection point of the adopter evolution (maximum of the first purchase unit sales) price fluctuations settle down. A very fast growing market is that of DVD players in the USA displayed in Fig. 4 [20]. This market has the highest supply growth rate $\alpha=0.65$ exhibiting a rapid increase of supplied units accompanied with a considerable price decline.

Displayed in Fig. 5 is the market evolution of Color TV sets in the USA [21]. The presented model is derived for the case that $\tilde{z}(t)$ does not suffer from supply constraints approaching $z_{max}=1$ with a constant supply growth rate $\alpha$ over the entire lifecycle $\Delta t$ (Eq. (16)). The empirical data suggest, however, that the supply evolution may have two separated growth periods with nearly constant $\alpha$. In this case the price evolution can be approximated by two logistic waves separated by time step $t_1$. This result can be interpreted as a rapid expansion of the production capacities after introduction of the good followed by a much slower output increase. Applying two successive logistic waves the price and hence the supply evolution can be fitted with higher accuracy. The data of the fit parameters $\alpha_k$, $z_{maxk}$, $C_{zk}$ with $k=0$ for the first and $k=1$ for the second growth period are summarized in Table 1.

A similar supply evolution can be found in the next two examples of durables. Shown in Fig. 6 is the market evolution of cell phones in the USA [20]. The data suggest a considerable increase of the output up to $t_1=1991$ accompanied by a rapid price decline. After this growth period the expansion of the capacities slowed down. A dichotomy of the supply evolution is also evident in the market evolution of VCR's in the USA [20] with two growth waves of nearly equal α.

**Table 1. Characteristic parameters of the studied examples, For simplicity we have set $\mu_m=0$**

| Parameter | B&W TV (UK) | B&W TV (USA) | CD player | DVD player | Colour TV | Cell phone | VCR |
|---|---|---|---|---|---|---|---|
| Figure | 1 | 2 | 3 | 4 | 5 | 6 | 7 |
| $t_0$ | 1946 | 1948 | 1983 | 1998 | 1954 | 1984 | 1978 |
| $\alpha$ [ $year^{-1}$] | 0.3 | 0.55 | 0.17 | 0.65 | 0.38 | 0.6 | 0.6 |
| $z_{max}$ | 1 | 1 | 1 | 1 | | | |
| $C_z$ | 8 | 10 | 6 | 4 | | | |
| $z_{max0}$ | | | | | 0.28 | 0.5 | 0.5 |
| $C_{z0}$ | | | | | 10 | 10 | 10 |
| $\alpha_0$ [ $year^{-1}$] | | | | | 0.38 | 0.6 | 0.6 |
| $t_1$ [ years] | | | | | 19 | 5 | 10 |
| $z_{max1}$ [14] | | | | | 0,72 | 0.5 | 0.5 |
| $C_{z1}$ | | | | | 6 | 10 | 150 |
| $\alpha_1$ [ $year^{-1}$] | | | | | 0.2 | 0.43 | 0.63 |
| $\mu_0$ [%] | 130 | 110 | 118 | 150 | 100 | 112 | 105 |
| $\mu_f$ [%] | 60 | 28.3 | 15 | 20 | 8 | 14 | 8 |
| H | 0.77 | 1.35 | 2.06 | 2 | 2.52 | 2 | 2.57 |
| A [ $year^{-1}$] | 0.0065 | 0.05 | 0.0085 | 0.007 | 0.0005 | 0.0015 | 0.0018 |
| B [ $year^{-1}$] | 0.4 | 0.4 | 0.45 | 1.2 | 0.4 | 0.55 | 0.64 |
| $n_0$ [%] | 90 | 90 | 76 | 70 | 83 | 55 | 82 |
| M in Mio | | | 0.85 | | | | 0.79 |
| $t_d$[ years] | 26 | | 17,8 | | | | 23.8 |
| $\theta$[ $year^{-1}$] | 0.16 | | | | | | |
| $n_{min}$ | 0 | | | | | | |
| a[ $year^{-1}$] | | | 0.3 | | | | 0.31 |
| a'[ $year^{-1}$] | | | 0.1 | | | | 0.65 |
| $\xi_{min}$ | | | 0 | | | | 0 |
| $\xi_{max}$ | | | 1.0 | | | | 0.33 |
| $C_\xi$. | | | 50 | | | | 120 |
| $\Delta t_q$[ years] | 20 | 14 | 18 | 10 | 25 | 15 | 16 |
| β | -1.03 | -1.29 | -1.25 | -0.57 | -1.02 | -1,1 | -1.03 |

[14] *The time dependence in Eq.(17) has to be replaced for the second wave by $t$-$t_1$.*

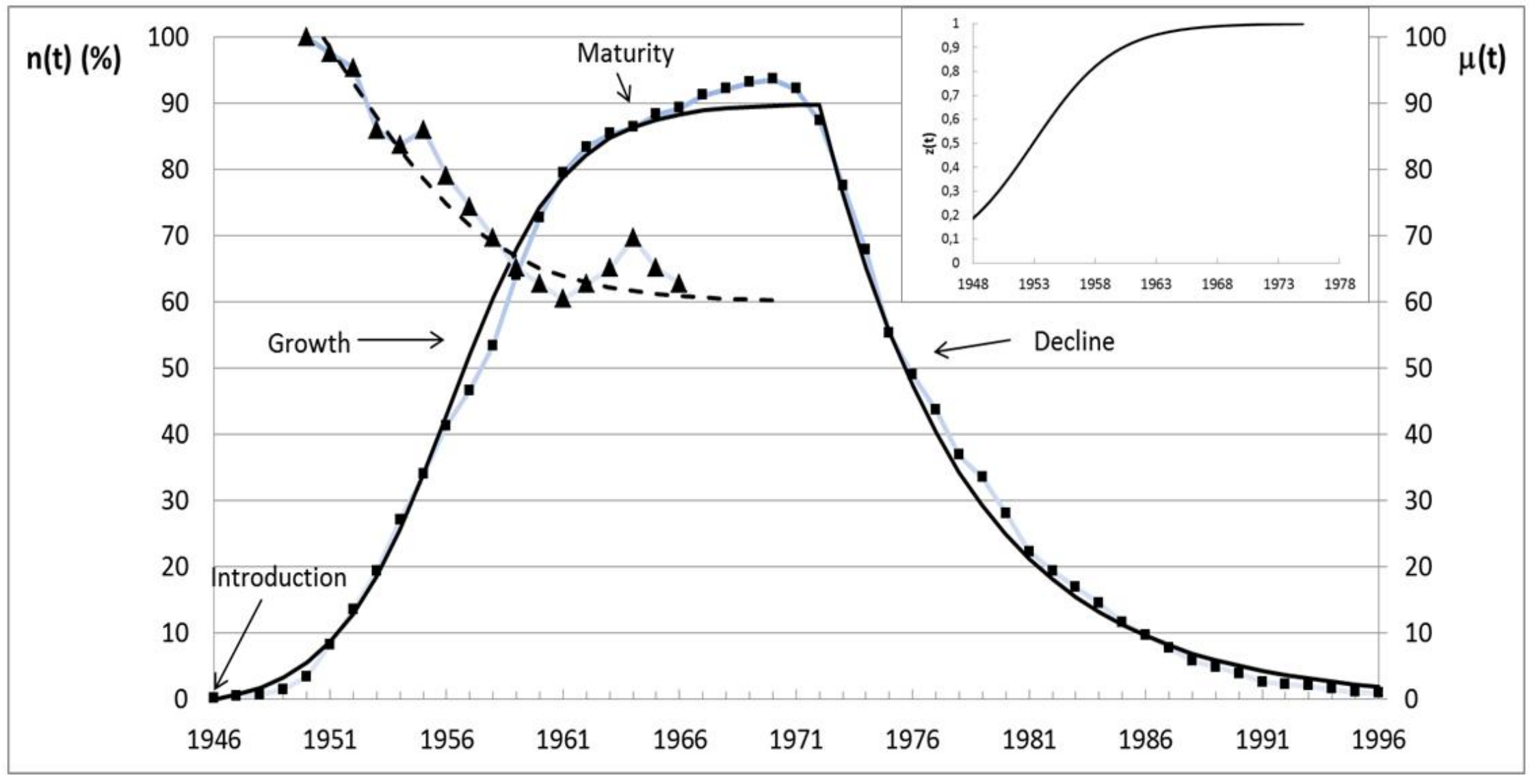

**Fig. 1. Evolution of the market penetration *n(t)* (squares) and mean price *μ(t)* (triangles) of Black & White TV sets in the UK [18,19]. The solid line (market penetration) and dashed line (mean price) are a fit with the parameters in Table 1. The insert shows the function $\tilde{z}(t)$ derived from the mean price evolution**

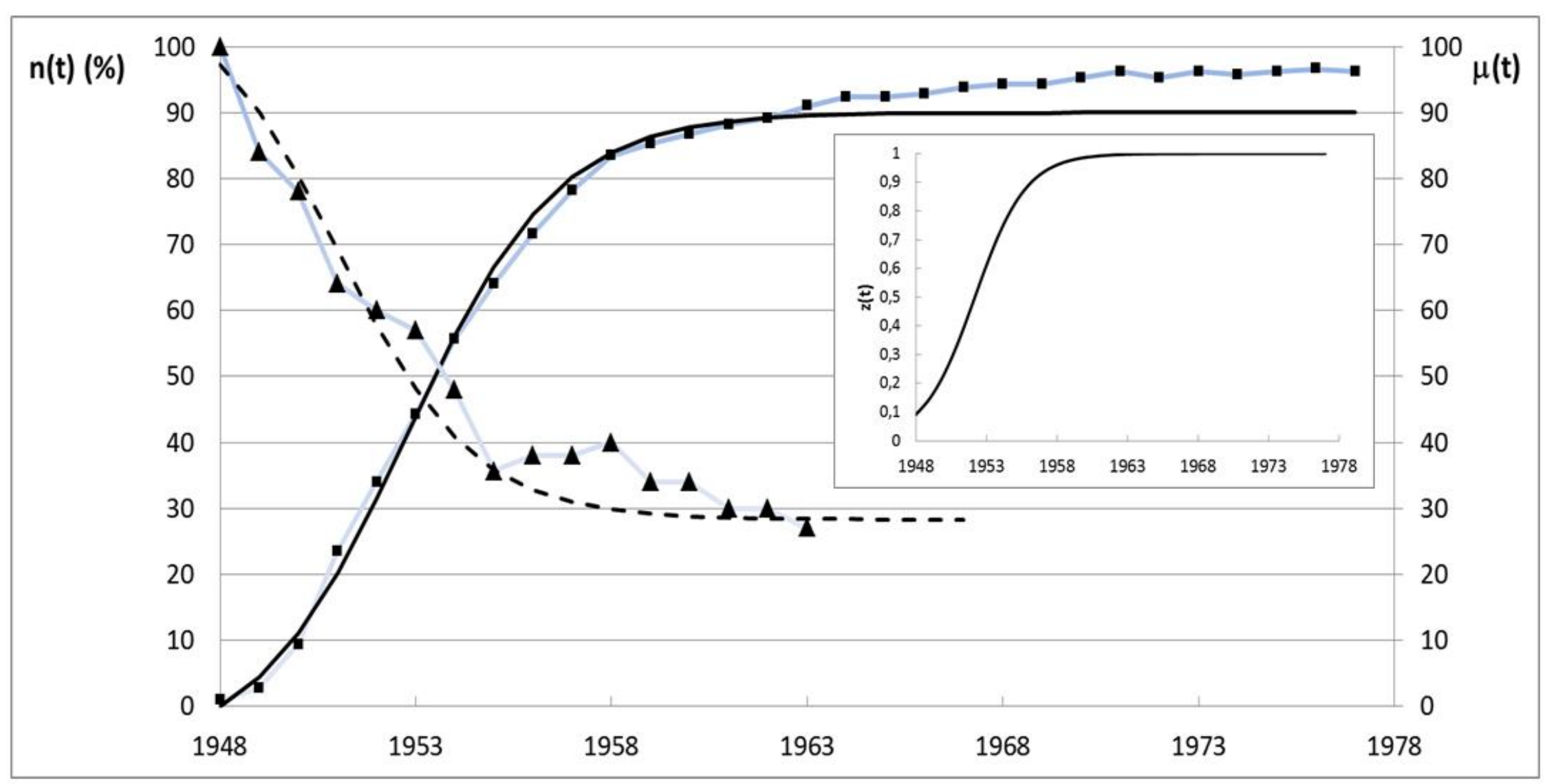

**Fig. 2. Evolution of the market penetration *n(t)* (squares) and mean price *μ(t)* (triangles) of Black & White TV sets in the USA [19]**

## 3.2 The Unit Sales Evolution

In addition to the market penetration and price evolution the model allows also an investigation of the evolution of the total unit sales. For two examples empirical sales data are available. Displayed in Fig. 8 are the total unit sales of VCR's in the USA (squares) together with a fit of Eq. (46) and Eq. (39) for first- and repurchase (solid line) with the parameters given in Table 1 [20]. The model suggests that the sales peak with its maximum around 1986 is due to first purchase caused by Bass diffusion. Thereafter the total unit sales increase due to repurchase. The decline of the unit sales starts in 2001 caused by the introduction of DVD recorders as a close substitute.

Another example of the PLC of a consumer durable is displayed in Fig. 9. It shows the rise and decline of the unit sales of CD-players in the USA (squares) and fit with the model equations (solid line) applying the parameters in Table 1 [20]. In difference to the VCR case the first sales peak caused by the diffusion process in not clearly evident. This is due to the slow penetration of this good. The time until it is replaced by a close substitute (MP3 player) is much shorter than for VCR's. The decline rate *a'* for CD-Players is, however, much smaller than for VCR's suggesting that CD-Players can be found in US households over long time horizons.

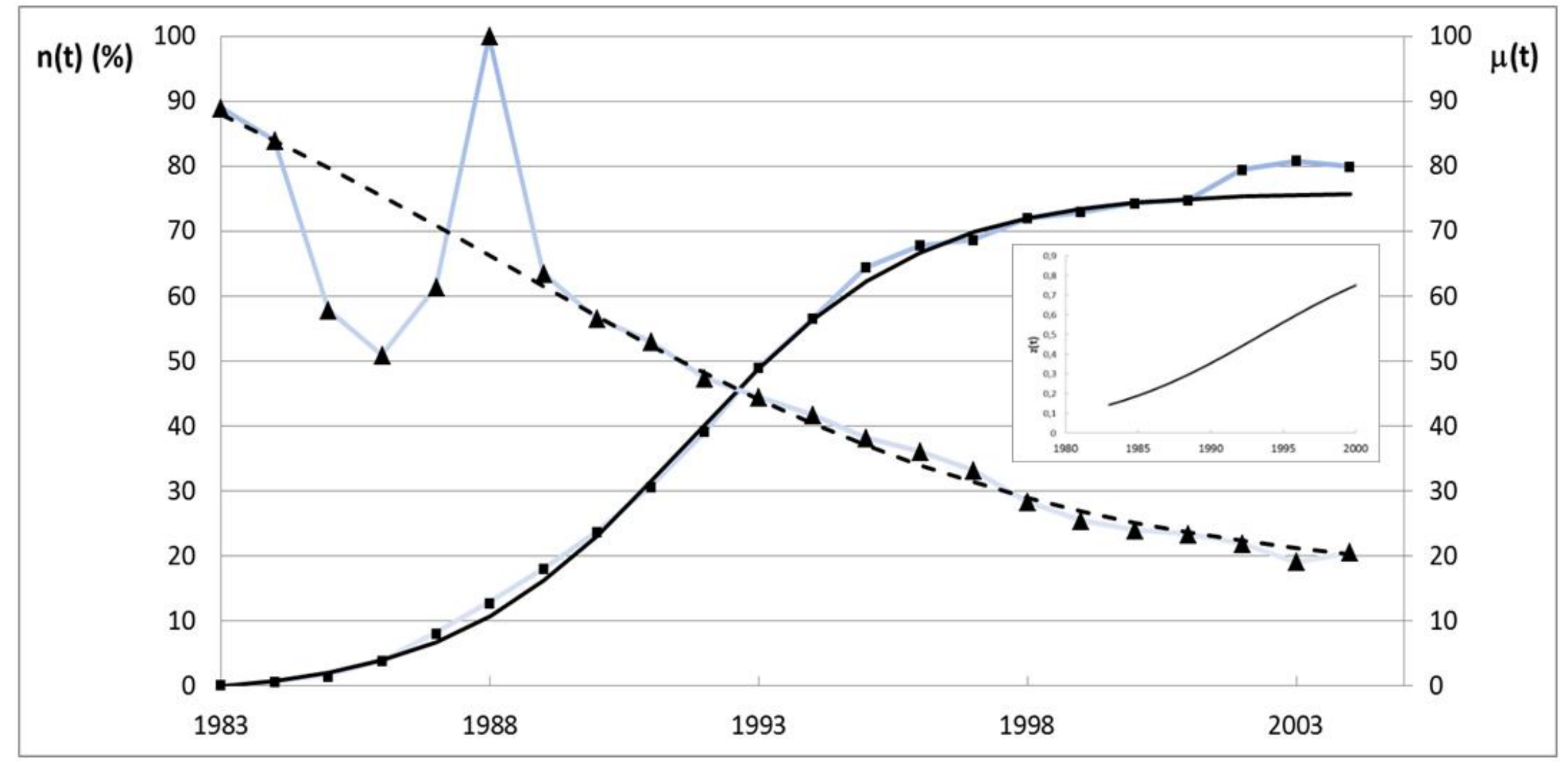

**Fig. 3. Evolution of the market penetration *n(t)* (squares) and mean price *μ(t)* (triangles) of CD players in the USA [20]**

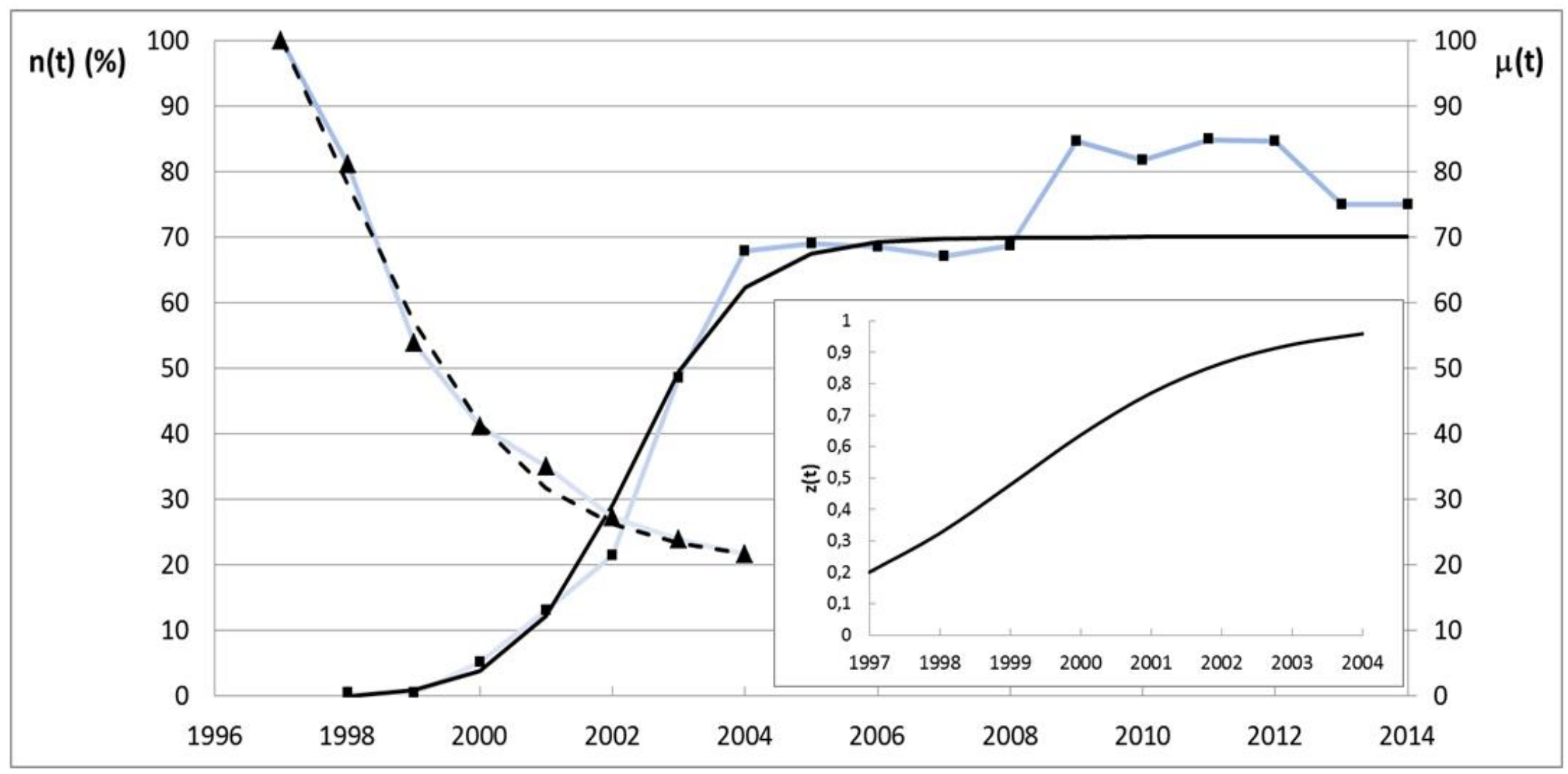

**Fig. 4. Evolution of the market penetration *n(t)* (squares) and mean price *μ(t)* (triangles) of DVD players in the USA [20]**

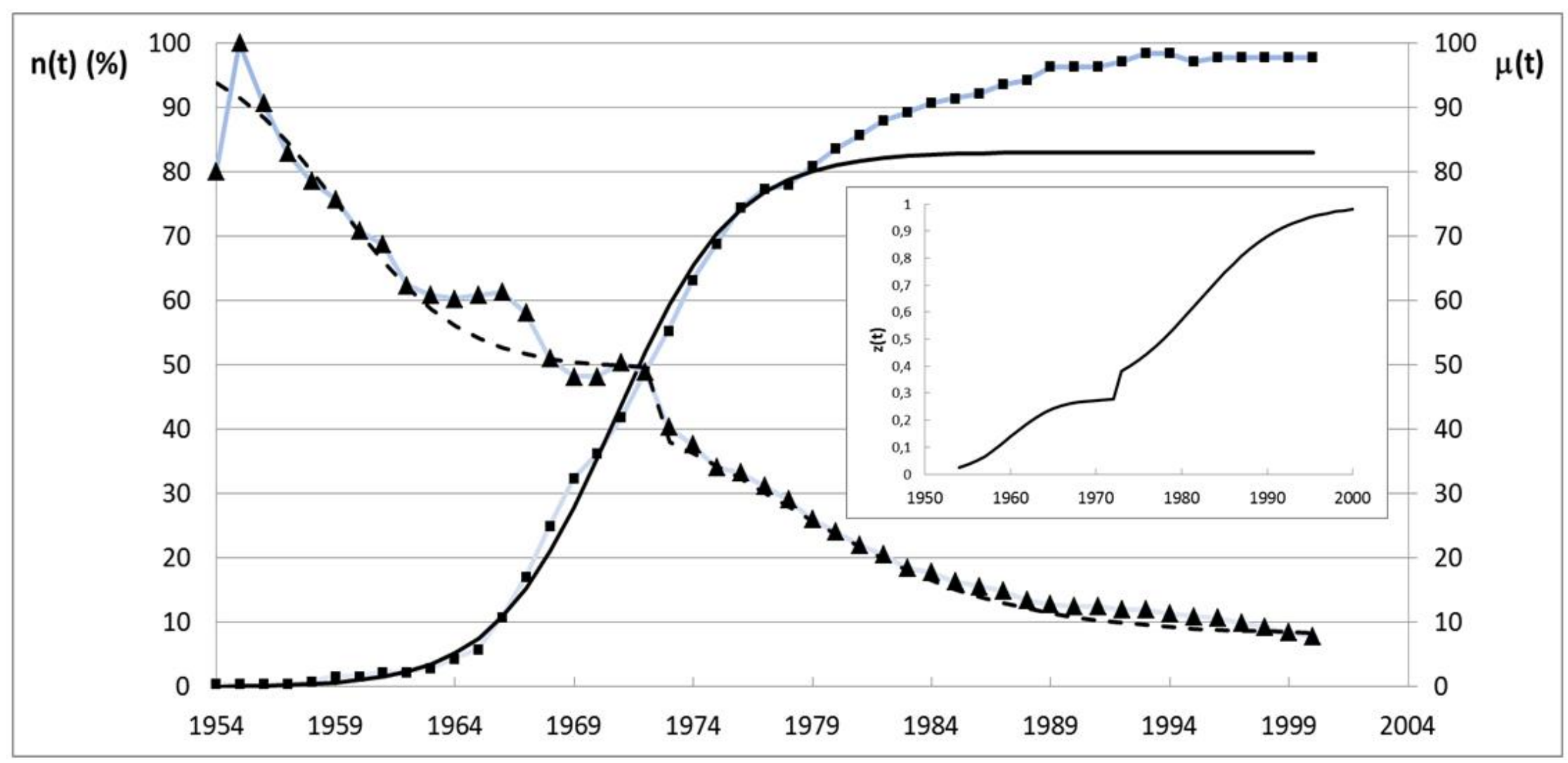

**Fig. 5. Evolution of the market penetration *n(t)* (squares) and mean price *μ(t)* (triangles) of Color TV sets in the USA [21]**

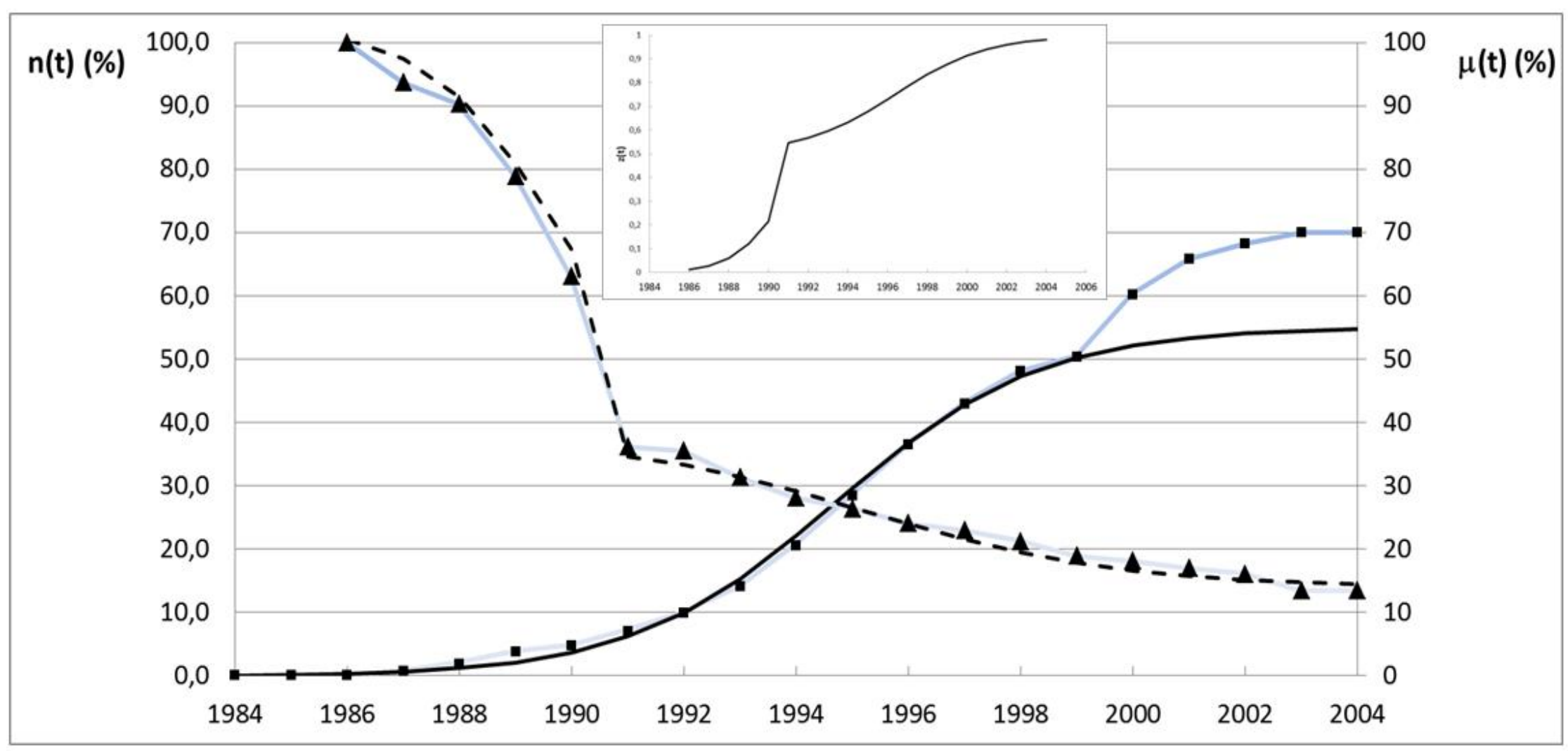

**Fig. 6. Evolution of the market penetration *n(t)* (squares) and mean price *μ(t)* (triangles) of cell phones in the USA [20]**

## 3.3 The Relation between Market Penetration and Supply

Eq.(55) predicts a linear relationship between the imitation parameter *B* of the Bass model as a measure of the diffusion velocity due to social contagion and the parameter *β* describing the impact of lost sales in the initial stages of the lifecycle. Taking advantage from the data in Table 1 the relation between both parameters is plotted in Fig. 10.[15] From a linear regression fit we get for the solid line displayed in Fig. 10 $B_0=1.7$ and $B_1=1.1$. Eq. (55) suggests, however, that both parameters should be equal. Since this is not exactly the case and due to the small coefficient of determination ($R^2=0.8$) of the regression fit we can conclude that the established theoretical considerations leading to Eq. (55) are qualitatively correct, but cannot verified quantitatively with this small number of examples.

[15] *In the case of two supply waves, the data of the first wave with index 0 were taken.*

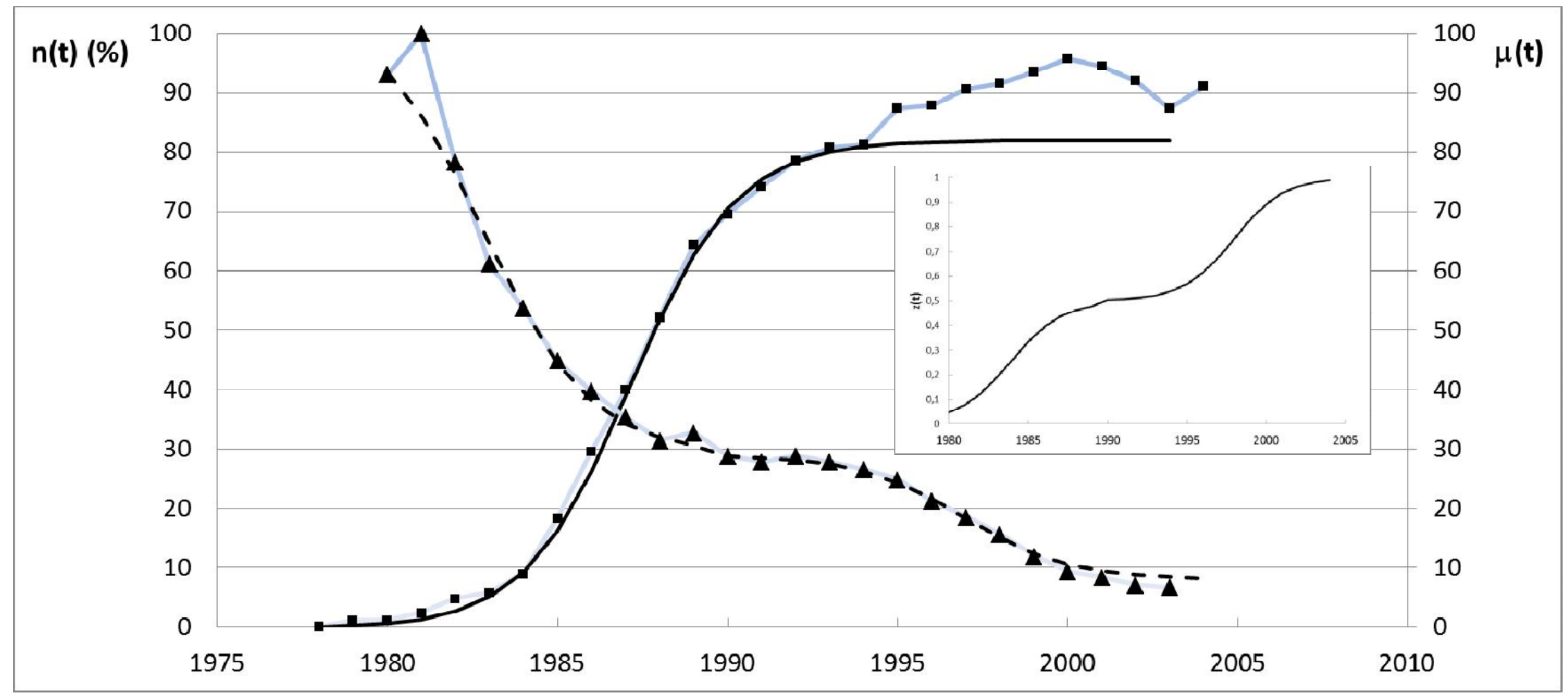

**Fig. 7. Evolution of the market penetration *n(t)* (squares) and mean price *μ(t)* (triangles) of VCR's in the USA [20]**

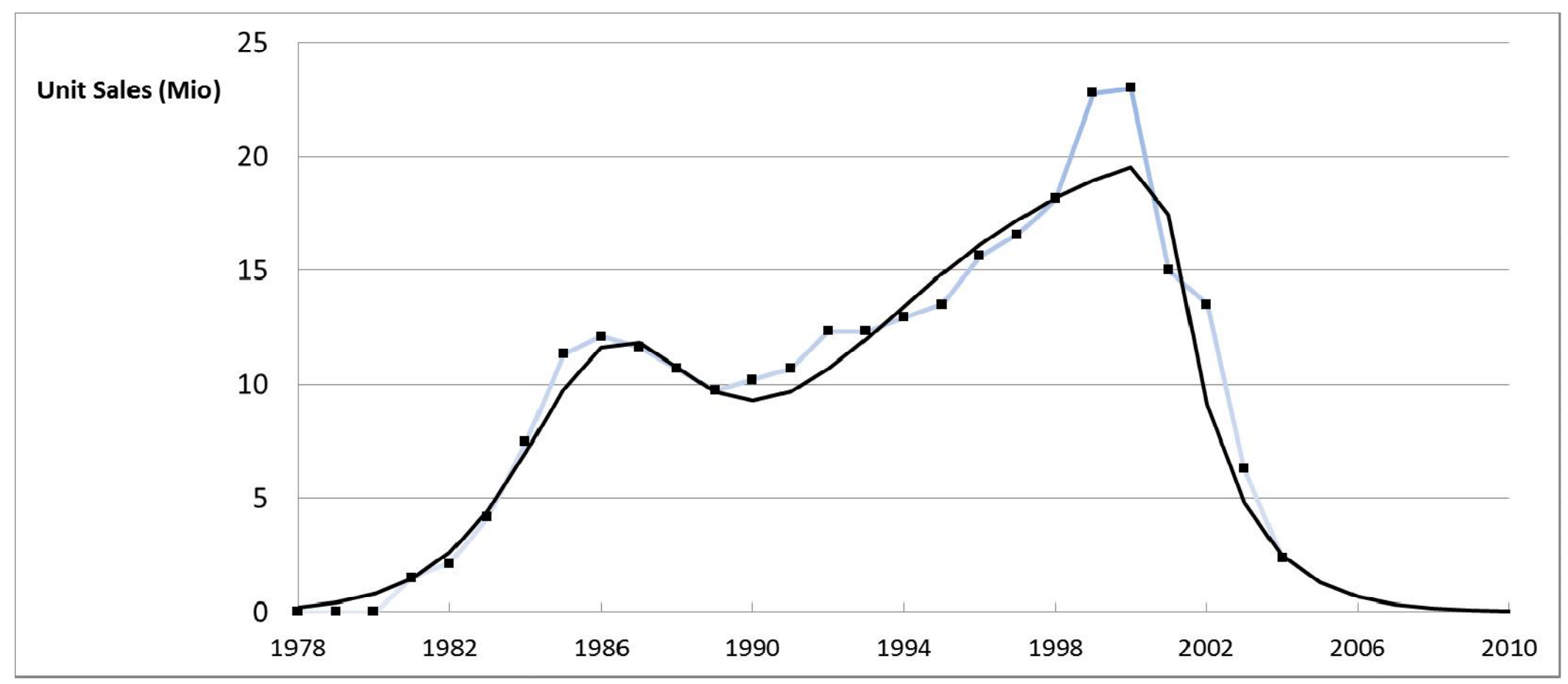

**Fig. 8. Evolution of the unit sales of VCR's (squares) in the USA [20]**

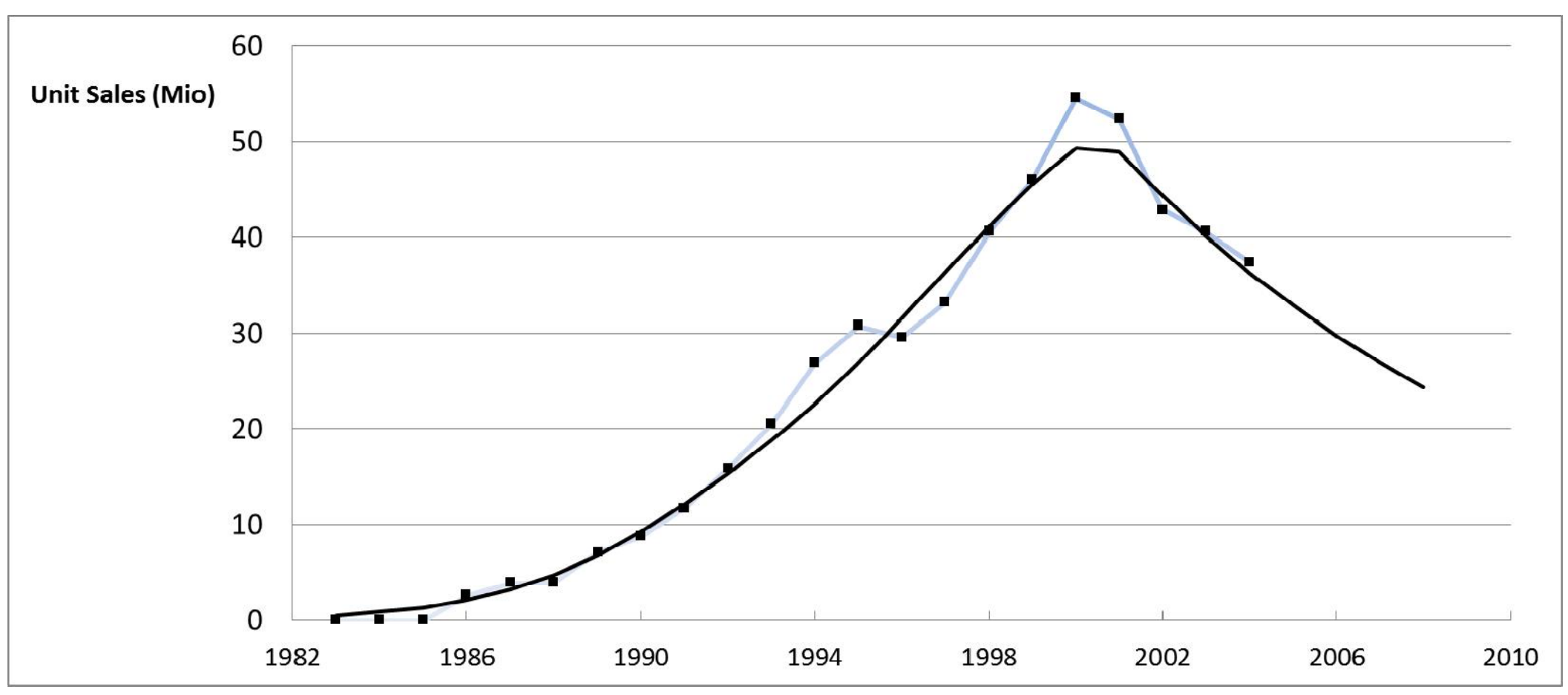

**Fig. 9. Evolution of the unit sales of CD players (squares) in the USA [20]**

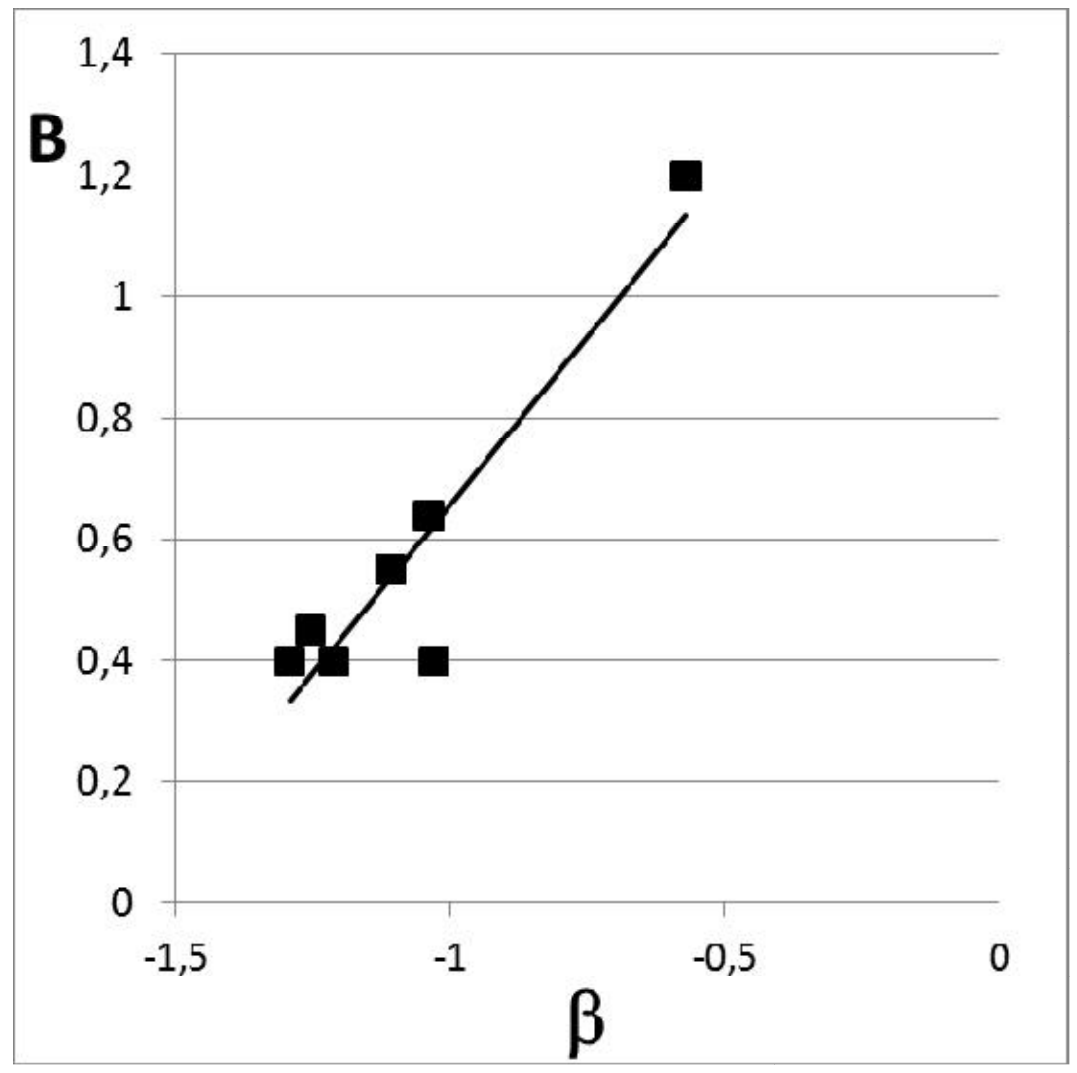

**Fig. 10. Displayed is the imitation parameter *B* (squares) as a function of *β* applying the data given in Table 1. The solid line is a linear regression function**

## 4. DISCUSSION

In this paper a dynamic microeconomic model of the PLC is established based on the dynamics of supply and demand. The PLC evolution is schematically displayed in Fig. 11 for the case of a homogeneous durable in a fast growing polypoly market. At introduction of a new durable the total number of available units $\widetilde{z}(t)$ (solid line) is usually small compared to the corresponding total number of demanded units $\widetilde{x}(t)$ (dashed line). As a consequence not all demands can be satisfied. [16]

The aggregated demand and supply curves *x(t,p)* and *z(t,p)* at introduction time step $t_0$ are indicated by solid lines in the left insert of Fig. 11. The model is based on the idea that sales events in a given price interval counted by price dependent unit sales rate *y(t,p)* are the result of a meeting process of demanded and supplied product units described by aggregated distribution functions *x(t,p)* and *z(t,p)*. The chance that demanded and supplied product units meet at *p* is determined by the overlap of these functions. This overlap has its maximum magnitude at the intersection point of *x(t,p)* with *z(t,p)*. The unit sales *y(t,p)* and also the price dispersion $P_y(p)$ have a maximum there. The intersection point corresponds to the mean price *μ(t)* of the good. In similarity to the neo-classic microeconomic approach the market price is therefore determined by the intersection of the supply and demand curves.[17] The resulting price dispersion $P_y(p)$ (dotted line) has for homogeneous goods the form of a Laplace distribution.

For a fast growing output (a supply market) the functions *x(t,p)* and *z(t,p)* are subject to time-dependent variations causing a shift of the mean price. The mean price evolution is governed in this model by a Walrus equation. It suggests that an excess increase of the number of demanded units increases and an excess increase of supplied units decreases the mean price in time. For the case of a fast supply growth the total number of available units $\widetilde{z}(t)$ increases in from of a logistic function until $\widetilde{z}(t) = z_{max}$, while the number of demanded units $\widetilde{x}(t)$ declines in the run of time, indicated in the upper graph of Fig. 11. As a result the mean price *μ(t)* declines governed by a logistic law approaching the floor price $\mu_f$ as displayed in the lower graph of Fig. 11 (For some durables the empirical data suggest that this evolution takes place in form of two logistic waves). Note that the price dispersion does not only shift to lower values in the PLC evolution, but because a price is strictly positive $P_y(p)$ must become a narrow peak near the floor price $\mu_f$ as schematically indicated in the right insert of Fig. 11 [13].

Accompanied with the supply and demand evolution the good spreads into the market described here by Bass diffusion. The spreading process determines the evolution of the market penetration *n(t)* and first purchase sales. Repurchase is taken into account here as proportional to *n(t)* and a logistic growth of the number of repurchased units per unit time and adopter.[18] The decline phase of a good starts with the introduction of a close substitute at time step $t_d$. It has an immediate impact on the repurchase behaviour of potential buyers causing an exponential decline of the total unit sales and the number of adopters with time. The total unit sales $\tilde{y}(t)$ are schematically displayed in the lower graph of Fig. 11 (dash-dot line) together with the market penetration *n(t)* (solid line). The characteristic stages of the lifecycle can be separated in this model by the main processes determining the adopter evolution also indicated in Fig. 11.

---

[16] *Note that in this dynamic microeconomic model the unrealistic assumptions of the neo-classic model that all demanded units are necessarily equal to all supplied units can be omitted. Not satisfied demands disappear and do not contribute to the unit sales (Eq. (11)).*

[17] *The demand and supply curves x(t,p) and z(t,p) are a consequence of the purchase process assuming a price minimizing propensity of potential buyers and a price maximizing propensity of suppliers [13]. A derivation of these functions from utility considerations of potential buyers or from the profit maximizing behaviour of the suppliers as in the neo-classic model is not required.*

[18] *Periodic oscillations in the repurchase process due to the initial growth period are neglected here (see [15]).*

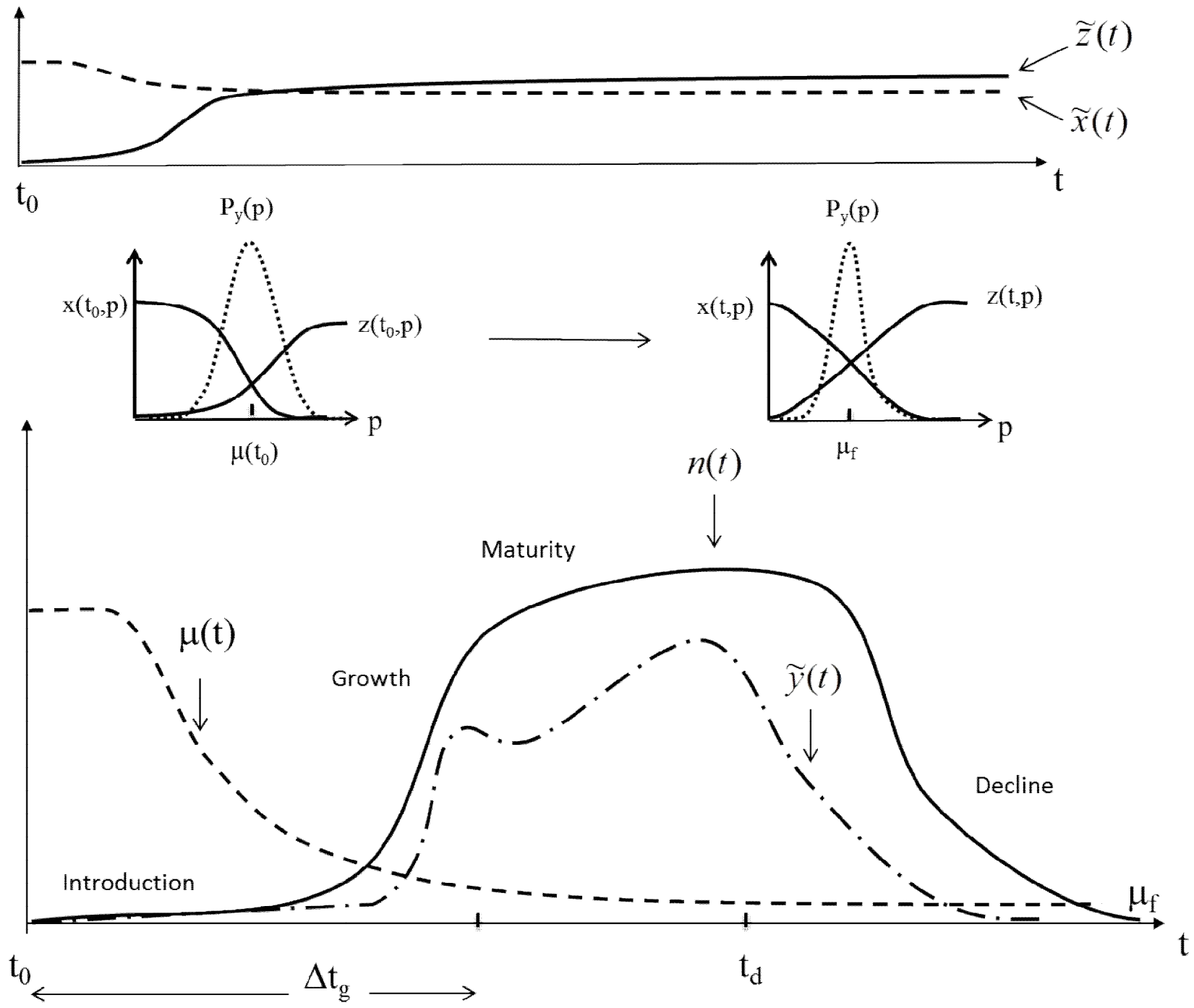

**Fig. 11. The product lifecycle of a durable homogenous good in polypoly supply markets suggested by the presented model**

Previous considerations suggested that the price evolution must have an impact on the diffusion process [22]. An additional price induced diffusion process is not further discussed here (see [15]). But the mean price expresses also the relation between supply and demand. Explicitly derived in this paper is that a supply shortage in the initial stages of the PLC may cause lost sales. This effect is the origin of a decelerated spreading of the good. The imitation parameter *B* of the Bass model has its maximum magnitude for a fast growing supply during the growth period $\Delta t_g$ (high supply growth parameter $\alpha$) and when the initial number of available units is high (small $C_z$). A comparison with empirical investigations shows that the theoretical considerations predicting a linear relationship between the imitation parameter *B* and a comprised parameter $\beta$ indicating the supply growth are qualitatively correct.

## 5. CONCLUSIONS

We can conclude that:

1. The PLC of homogeneous durable goods in polypoly markets is determined by the complex dynamics of demand and supply. This dynamics generates a (Laplacian) price dispersion while the evolution of the mean price is governed by a Walrus equation. The demand side is characterized by the evolution first- and repurchase events and the supply side by the output growth.
2. First purchase is related to a spreading process of the good into the market where the market penetration is governed by a conservation relation. The initial stages of the PLC can be well described by the Bass model suggesting a fast and a slow spreading wave where the fast is mediated by mass media and the slow is due social contagion.
3. Repurchase is caused by the finite lifetime of a good and multiple purchase events. It

can be modelled as a logistic growth of repurchase events.

4. The introduction of a close substitute changes the repurchase behaviour of potential buyers. It induces an exponential decline of the repurchase sales and market penetration.
5. Supply constraints in the initial stages of the PLC cause a retarded spreading of the good into the market. The imitation parameter of the Bass model decreases with a slow output growth and a small number of available units at introduction of the good.

## COMPETING INTERESTS

Author has declared that no competing interests exist.